\documentstyle[12pt]{article}
\newcommand{\be}[1]{
\begin{eqnarray}\label{#1}}
\newcommand{\ee}{\end{eqnarray}}
\newcommand{\ci}[1]{\cite{#1}}
\newcommand{\re}[1]{(\ref{#1})}
\newcommand\pbar{\bar{\psi}}
\newcommand\p{\psi }
\newcommand\prl{\partial}
\def\Gtilde{\tilde{G}}
\def\xslash{\rlap/{\mkern-1mu x}}
\newcommand{\pert}{{\cal D}}

  \def\beq{\begin{equation}}
  \def\eeq{\end{equation}}
  \def\beqr{\begin{eqnarray}}
  \def\eeqr{\end{eqnarray}}
  
\begin{document}
\renewcommand{\thefootnote}{\fnsymbol{footnote}}
\begin{flushright}
\begin{tabular}{l}
\\
TPR-00-11\\
\end{tabular}
\end{flushright}
\begin{center}
{\bf\Large Wandzura-Wilczek approximation
for the twist-3 DVCS amplitude}

\vspace{0.5cm}
N. Kivel$^{a,b}$,  M.V. Polyakov$^{b,c}$,
A. Sch\"afer$^{a}$
O.V. Teryaev$^{a,d}$
\begin{center}
{\em
$^a$ Institut f\"ur Theoretische Physik, Universit\"at Regensburg \\
D-93040 Regensburg, Germany \\
$^b$Petersburg Nuclear Physics Institute,
188350, Gatchina, Russia\\
$^c$ Institut f\"ur Theoretische Physik II, Ruhr-Universit\"at Bochum,\\
D-44780 Bochum, Germany \\
$^d$ Bogoliubov Laboratory of Theoretical Physics, JINR,
141980, Dubna, Russia}
\end{center}
 \end{center}
 \vspace{1.5cm}

 \begin{abstract}
We present a derivation of  Wandzura-Wilczek (WW) like relations for
skewed parton distributions. It is demonstrated  for photon-pion scattering
that the skewed twist-3 parton
distributions contributing to the DVCS amplitude
have discontinuities at the points $x=\pm\xi$ in the WW
approximation. This may lead to a violation of
factorisation for the twist-3 DVCS amplitude with transverse
polarization of the virtual photon.
We show, however, that the contribution of the divergencies
to the scattering of a transversely polarized virtual photon
affects DVCS observables only at order $1/Q^2$ and can be neglected
at  twist-3 accuracy.
For the scattering of a
longitudinally polarized photon the twist-3 amplitude is free of such
divergencies.
 \end{abstract}

\section*{\normalsize \bf Introduction}

Deeply Virtual Compton Scattering (DVCS) \cite{DVCS1,DVCS2} is the
cleanest hard process which is
sensitive to the Skewed Parton Distributions (SPD) and has been
the subject of extensive theoretical investigations for a few years.
First experimental data became recently available (see e.g. \cite{exp1,exp2})
and much more data are expected from JLAB, DESY, and CERN in the near future.
It was demonstrated  \cite{Rad97,Ji98,Col99} that in leading
order  ($1/Q^0$) the DVCS amplitude  factorizes.
However, as the typical $Q^2$ are by no means large, studies
of the power corrections to the DVCS amplitude are very important.

Recently the DVCS amplitude  was computed
\cite{Anikin,Penttinen,BM} including the terms of order $O(1/Q)$.
The inclusion of such terms is mandatory to conserve the
electromagnetic gauge invariance of the DVCS amplitude at this
order of $1/Q$. Moreover, they provide the leading contribution to
some spin asymmetries. To this order the amplitude depends on a
set of new skewed parton distributions. Recently it was shown in
ref.~\cite{BM} that in the so-called Wandzura-Wilczek (WW)
approximation these new functions can be expressed in terms of
twist-2 SPD's. WW relations for SPD's were also discussed in
ref.~\cite{BR}, however the authors of this paper did not take
into account operators which are total derivatives. Obviously
these operators are crucial for the description of non-forward
matrix elements.

In this note we give a derivation of the WW-like relations
which  is technically slightly different from the approach
used in \ci{BM},  see the
Appendix for details. Then we demonstrate for the case of
DVCS on the pion the new SPD's in WW approximation
generically posses discontinuities at the
points $x=\pm \xi$. This leads to a formally divergent expression
for the DVCS amplitude, which however contributes to the DVCS observables
at the accuracy $1/Q^2$ and hence beyond our accuracy.

We also observed that in the twist-3 DVCS amplitude\footnote{We use
the notion of 'twist-3 amplitude' as synonymous to  'amplitude
of order $1/Q$'}
with {\it
longitudinally} polarized virtual photons the
divergencies mentioned above are cancelled.
We discuss the
theoretical implications of this phenomenon and its possible experimental
verification.

\section*{\normalsize \bf DVCS on the pion}

The expression for the DVCS amplitude
on the pion as obtained in \cite{Anikin}
and reproduced in refs.~\cite{Penttinen,BM} has the form:
\be{T}
T^{\mu\nu} &=&
-\frac{1}{2P\cdot Q}\int dx
\Biggl(
\frac{1}{x-\xi+i\epsilon} + \frac{1}{x+\xi-i\epsilon}
\Biggr)
\Biggl[
H_1(x,\xi)
\Biggl(
-2\xi P^{\mu}P^{\nu}  -
P^{\mu}Q^{\nu} - P^{\nu}Q^{\mu}
\Biggr.
\nonumber\\
\Biggl.
&+&g^{\mu\nu}(P\cdot Q) - \frac{1}{2}P^{\mu}\Delta^{\nu}_{\perp} +
\frac{1}{2}P^{\nu}\Delta^{\mu}_{\perp}
\Biggr)
\nonumber\\
&-&\bigl[ H_3(x,\xi)+\frac \xi x H_A(x,\xi)\bigr]
\Delta^{\nu}_\perp
\Biggl(
3\xi P^{\mu} + Q^{\mu}
\Biggr)
\nonumber\\
&-&\bigl[ H_3(x,\xi)-\frac \xi x H_A(x,\xi)\bigr]
\Delta^{\mu}_{\perp}\Biggl(
\xi P^{\nu} + Q^{\nu} \Biggr) \Biggr]\, ,
\ee
where $P=(p+p')/2, Q=(q+q')/2,
\Delta=q-q'=p'-p = -2 \xi P +\Delta_\perp$, while $p,p',q,q'$ are the
initial and final momenta of pion and photon, respectively.
The three terms in the square brackets are individually gauge invariant up
to order $\Delta_{\perp}^2$, i.e. $q'_{\nu}T^{\mu\nu}=O(\Delta_{\perp}^2)=
q_{\mu}T^{\mu\nu}$.
The second and third term contain the new
twist-3 contributions to the DVCS amplitude. As $P$ and $Q$ are longitudinal
and  $\Delta_{\perp}$ transverse the second corresponds to
longitudinally polarized virtual
photon and  third to transverse polarization.

The amplitude depends on new SPD's defined as\footnote{
Note that the function $H_A$ in ref.~\cite{Anikin} is defined with opposite
sign}
\be{Parm}
\int \frac{d\lambda}{2\pi} e^{i\lambda x (\bar P n)}
\langle p'| \bar \psi\biggl(-\frac{\lambda}{2}n
\biggr) \gamma^\mu
\psi\biggl(\frac{\lambda}{2}n\biggr)|p\rangle&=&
P^\mu H_1(x,\xi) + \Delta^\mu_\perp H_3(x,\xi)\, , \\
\int \frac{d\lambda}{2\pi} e^{i\lambda x (\bar P n)}
\langle p'| \bar \psi\biggl(-\frac{\lambda}{2}n
\biggr) \gamma^\mu \gamma_5
\psi\biggl(\frac{\lambda}{2}n\biggr)|p\rangle&=&
i\varepsilon_{\mu \alpha \beta \delta}\Delta^{\alpha}P^{\beta}
n^{\delta}H_A(x,\xi) \, .
\ee
Here the light-cone vector $n$ is normalized as $n\cdot P=1$

Recently in ref.~\cite{BM} it was suggested to use the
Wandzura-Wilczek approximation to express functions
like $H_3$ in terms of twist-2 function $H_1$.
This approximation gives the kinematical part of twist 3,
while the genuine, dynamical, higher twist is neglected.
If one adopts the operator definition of twist,
this
approximation, because of neglecting the operators containing the gluon
field strength, corresponds to twist 2.
In the Appendix we give alternative derivation of such relations
and in the next section we show that the WW approximation
generically leads to formally divergent expression for the DVCS
amplitude. We also specify the twist-3 helicity amplitudes which
are free of such divergencies.

\section*{\normalsize \bf Discontinuities of
the twist-3 distributions in WW approximation}

In this section we study the properties of the twist-3
distributions in the WW approximation. We shall see that generically
they posses discontinuities at the points $x=\pm \xi$. We shall illustrate
this by the example of  $H_3(x,\xi)$.

Using the general expression (\ref{reslt}) discussed in the Appendix
it is easy to obtain the
function $H_3(x,\xi)$ in WW approximation.
It can be expressed in terms
of the twist-2 SPD $H_1(x,\xi)$ as follows (c.f. \cite{BM}):

\be{ww3}
\nonumber
 H_3^{WW}(x,\xi)&=&-\frac 14 \Biggl\{
 \theta(x>\xi) \int_x^1\frac{du}{u-\xi}
\biggl[ \frac{\partial H_1(u,\xi)}{\partial
\xi}+ \frac{\partial
H_1(u,\xi)}{\partial
u}
\biggr] \\
\nonumber
&-&\theta(x<\xi) \int_{-1}^x \frac{du}{u-\xi}
\biggl[ \frac{
\partial H_1(u,\xi)}{
\partial \xi}+ \frac{
\partial H_1(u,\xi)}{\partial u}
\biggr]\\
&+& \theta(x>-\xi) \int_x^1\frac{du}{u+\xi}
\biggl[ \frac{
\partial H_1(u,\xi)}{
\partial
\xi}- \frac{
\partial
H_1(u,\xi)}{
\partial u}
\biggr] \\
\nonumber
&-&\theta(x<-\xi) \int_{-1}^x \frac{du}{u+\xi}
\biggl[ \frac{
\partial H_1(u,\xi)}{
\partial
\xi}- \frac{
\partial
H_1(u,\xi)}{
\partial
u}
\biggr] \Biggl\} \,.
\ee

Note that in the limit $\xi=0$ this reduces to an expression,
which is similar to
the standard WW approximation for the transverse spin structure
function $g_T$.
Note that the specific form of (4)
guarantees the correct symmetry properties of
$H_3$ \cite{Anikin}, which is a sensitive check of our result.
Another interesting  limit  to discuss is $\xi=1$.
In this limit  our expression is similar
to the corresponding relation for the light cone amplitude for
vector mesons (which is the closest case worked out in literature, namely
 by Ball and Braun \cite{BB96}).
The relation (\ref{ww3}) also allow to generalize the WW relation
for vector mesons distribution amplitudes obtained
in \cite{BB96} to the case of a meson of arbitrary spin.
This can be done with help of
crossing relations between pion SPD's and
distribution amplitudes of resonances \cite{MVP98}.

{}From eq.~(\ref{ww3}) one can derive the following relations
for the Mellin moments of $H_1$ and $H_3$:
\be{mom0}
\int_{-1}^1 dx H_3(x,\xi)&=&-\frac 12 \ \frac{d}{d\xi}
\int_{-1}^1 dx
H_1(x,\xi)=0,
\ee
where the last equation follows from the polynomiality condition
for SPDs \cite{DVCS2}.
This relation is an analog of the Burkhardt-Cottingham sum rule,
it is valid beyond WW approximation.
This particular moment, like the other even moments,
corresponds to the non-singlet part of the SPDs, while the
contribution of the singlet part is zero for the trivial reason
that the pion SPD antisymmetric in $x$.
Another interesting relation,
which is an analog of the Efremov-Leader-Teryaev sum rule \cite{ELT},
can be obtained
for the second Mellin moments of the SPD's, corresponding to the singlet
part, appearing in  DVCS:
\be{mom1}
\int_{-1}^1 dx x\ H_3(x,\xi)&=&-\frac14 \ \frac{d}{d\xi}
\int_{-1}^1 dx x\
 H_1(x,\xi)=\frac 12\xi M_2,
\ee
where $M_2$ is a momentum fraction carried by the quark in the pion.
To obtain the last equality we used the generalized momentum sum rule
for SPDs at $\Delta^2=0$
as derived in \cite{PW}. Let us stress that this relation is
valid beyond the WW approximation. Its generalizations to the nucleon
case was recently discussed in \cite{Penttinen}.

With the help of expression (\ref{ww3}) we can easily compute the behaviour
of $H_3$ near the points $x=\pm \xi$. Let us consider the
difference of left and right limits of the function $H_3$ at
$x\to\pm \xi$, the result is:
\be{jump}
\lim_{\delta\to 0}
\biggl\{H_3^{WW}(\xi+\delta,\xi)- H_3^{WW} (\xi-\delta,\xi)
\biggr\} = -\frac 14
vp \int_{-1}^{1} \frac{du}{u-\xi}\
\biggl[ \frac{\prl H_1(u,\xi)}{\prl\xi}+ \frac{\prl H_1(u,\xi)}{\prl u}
\biggr]\, ,\\
\lim_{\delta\to 0}
\biggl\{ H_3^{WW} (-\xi+\delta,\xi)- H_3^{WW}(-\xi-\delta,\xi)
\biggr\} = -\frac 14
vp \int_{-1}^{1} \frac{du}{u+\xi}\
\biggl[ \frac{\prl H_1(u,\xi)}{\prl\xi}- \frac{\prl H_1(u,\xi)}{\prl u}
\biggr]\, ,
\ee
where $vp \int$ denotes a principle value integral ({\it valeur principal}).
From these expressions we see that for a wide class
of functional forms for $H_1$ the corresponding function $H_3$
exhibits discontinuities at the points $x=\pm \xi$. Even for
smooth functions $H_1$ the discontinuities are non-zero.
To clarify this point, it is instructive to
integrate by part, imposing the natural condition $H_1(1,\xi)=
H_1(-1,\xi)=0$. One gets:
\be{j2}
\lim_{\delta\to 0}
\biggl\{H_3^{WW}(\xi+\delta,\xi)-H_3^{WW} (\xi-\delta,\xi)
\biggr\} &=& -\frac 14
\frac{d}{d \xi}\ vp \int_{-1}^{1} \frac{du H_1(u,\xi)}{u-\xi} , \\
\lim_{\delta\to 0}
\biggl\{H_3^{WW}(-\xi+\delta,\xi)-H_3^{WW} (-\xi-\delta,\xi)
\biggr\} &=& -\frac 14
\frac{d}{d \xi}\ vp \int_{-1}^{1} \frac{du H_1(u,\xi)}{u+\xi}\, .
\ee
The sum of the two jumps which arises in the convolution integral for
the amplitude can be expressed in terms of the
real part of the twist-2 DVCS amplitude as follows:
\be{j3}
\nonumber
\lim_{\delta\to 0}
\biggl\{H_3^{WW}(\xi+\delta,\xi) - H_3^{WW} (\xi-\delta,\xi)
+H_3^{WW}(-\xi+\delta,\xi)
-H_3^{WW} (-\xi-\delta,\xi)
\biggr\} \nonumber \\=-\frac 14
\frac{d}{d \xi}\ vp \int_{-1}^{1} du H_1(u,\xi)
\biggl\{\frac{1}
{u+\xi}+\frac{1}{u-\xi}
\biggr\}.
\ee
The latter expression is nothing else than the derivative of the real
part of twist two DVCS amplitude. As DVCS is an observable physical
process (and as already been observed) the integral on the right hand side
of (\ref{j3}) cannot be zero and because the  physics has to show
a non-trivial dependence on the skewedness parameter it  cannot be
independent of $\xi$.
This shows that the r.h.s. of (\ref{j3}) cannot vanish and therefore
$H_3$ really has to show discontinuities.

The expression for the WW- part $H_A^{WW}(x,\xi)$ of the function
$H_A(x,\xi)$ is slightly different:
\be{wwHA}
\nonumber
  H_A^{WW} (x,\xi)&=&\frac 1{4\xi} \Biggl\{
 \theta(x>\xi) \int_x^1\frac{du}{u-\xi}
\biggl[
u \frac{\prl H_1(u,\xi)}{\prl u}+\xi \frac{\prl H_1(u,\xi)}{\prl \xi}
\biggr] \\
\nonumber
&-&\theta(x<\xi) \int_{-1}^x \frac{du}{u-\xi}
\biggl[
u \frac{\prl H_1(u,\xi)}{\prl u}+\xi \frac{\prl H_1(u,\xi)}{\prl \xi}
\biggr]\\
&-& \theta(x>-\xi) \int_x^1\frac{du}{u+\xi}
\biggl[
u \frac{\prl H_1(u,\xi)}{\prl u}+\xi \frac{\prl H_1(u,\xi)}{\prl \xi}
\biggr] \\
\nonumber
&+&\theta(x<-\xi) \int_{-1}^x \frac{du}{u+\xi}
\biggl[
u \frac{\prl H_1(u,\xi)}{\prl u}+\xi \frac{\prl H_1(u,\xi)}{\prl \xi}
\biggr] \Biggl\} \,.
\ee
It can be checked that the function $H_A^{WW}(x,\xi)$ also posses
discontinuities at the points $x=\pm \xi$. The expression for these
discontinuities has the form:
\be{jumpA}
\lim_{\delta\to 0}
\biggl\{H_A^{WW}(\pm \xi+\delta,\xi)-H_A^{WW}
(\pm \xi-\delta,\xi)
\biggr\} = \pm \frac 14
\frac{d}{d \xi}\ vp \int_{-1}^{1} \frac{du H_1(u,\xi)}{u \mp \xi} .
\ee

Note that the difference of these discontinuities (which is the
combination
entering the amplitude (\ref{T})) coincides with (\ref{j3}).
The presence of discontinuities in the functions $H_3$ and $H_A$
exactly at the points $x=\pm \xi$ leads to a formally divergent
result for the amplitude at the order $O(1/Q)$, see expression
eq.~(\ref{T}). This indicates a violation of
factorisation in order $O(1/Q)$ for the DVCS amplitude in WW approximation.
The considered violation
disappears for virtual final photon, as the pole of the
quark propagator no more occurs at the point $x=\xi$.
Thus the production of Drell-Yan pairs does not pose any  problems.
Experimental studies comparing real
and virtual photon production thus  provide
an excellent  opportunity to check factorisation.

It is important to realize that the formal divergencies of the amplitude
in the WW approximation are canceled for certain combinations of
helicity amplitudes because the jumps in $H_3$ and $H_A$ are related to
each other, see eqs.~(\ref{j3},\ref{jumpA}).
Since the divergencies occur only at the point $x=\pm\xi$
it is sufficient that the jumps cancel at this specific value to
save factorisability.
Specifically there is no problem for
the amplitudes with {\it longitudinal} polarization of the virtual
photon, because  the corresponding amplitude is proportional
to the following combination of the functions $H_3$ and $H_A$:
\be{Tlong}
\varepsilon_L^\mu T_{\mu\nu}\propto
\Delta_{\perp\nu}\int dx
\Biggl(
\frac{1}{x-\xi+i\epsilon} + \frac{1}{x+\xi-i\epsilon}
\Biggr) \biggl[ H_3(x,\xi)+\frac{\xi}{x} H_A(x,\xi)
\biggr].
\ee
Thus the discontinuities generate divergencies only for the twist-3
DVCS amplitude with {\it transversely} polarized virtual photon.
The contribution of the ``problematic"
part of the twist-3 DVCS amplitude (corresponding to scattering
of the transversely polarized virtual photons)
to the differential cross section is suppressed
after the contraction with the polarization vector of the emitted
real photon
by two powers of the hard scale, i.e. by $1/Q^2$,  relative to the
leading order result and do not contribute to observables in order
O($1/Q$). The DVCS differential cross section to this accuracy
gets contributions only from the longitudinal part of the twist-3
amplitude which is free from divergencies.

The physical reason of these peculiarities for
the transverse photon case is
rather similar to the origin of the famous Callan-Gross relation.
The dangerous pole of the quark propagator,
which in combination with jumps lead to the
 violation of factorisation,  corresponds
to an on-shell quark, which may absorb only  transverse photons.
We therefore expect that
this situation should persist also in
the case of  nucleon and deuteron targets.
This  would be interesting to check  by an explicit calculations.

The considered relationship between twist-2 and twist-3 terms
can also be discussed with regard to
electromagnetic gauge invariance.
The leading twist-2 term is gauge invariant
only if one neglects the transverse component
of the momentum transfer $\Delta_\perp$. In $O(\Delta_\perp)$ accuracy
it requires the consideration of a quark-gluon diagram, whose contribution
by use of the equation of motions is expressed through the
twist-2 SPD $H_1$ and the new twist three SPDs $H_3, H_A$.
While the $H_1$ term combines with the leading result to
a gauge-invariant expression (see two first lines in
eq.~(\ref{T})), as anticipated in
ref.~\cite{GuichonVander}, the other terms provide the additional
contribution which violates factorisation in the WW approximation for
transverse polarization of the virtual photon.

Generally, the appearance of the jumps just at the points $x=|\xi|$
is not unnatural from the physical point of view, as this is just a
transition point between the regions, where SPD has quite different
physical meaning \cite{Rad97}, accommodated, in particular, in the
two-component model of SPDs \cite{PW}.

As  final remark we note that preliminary
estimates of the function
$H_3$ at a low normalization point
in the instanton model give a function without discontinuities
and hence indicates that the WW approximation for SPD's
is not valid within this specific  model.

\section*{\normalsize \bf Conclusions}

We demonstrated by explicit calculations that Wandzura-Wilczek
like relations for SPD's entering the
description of the DVCS amplitude
at order $O(1/Q)$ lead to a violation of factorisation
for the twist-3 DVCS amplitude with transversely polarized virtual
photons. However, one can easily see that the dangerous divergencies
do not contribute to DVCS  observables at the order $1/Q$ but
at the order $1/Q^2$. Therefore these divergencies affect only
twist-4 corrections which are beyond the scope of the present paper.
One cannot exclude that the kinematical contributions of twist 4 will cancel
the considered divergencies. This promising opportunity is suggested
by the paper \cite{RW}, in which a 
part of the $1/Q^2$ term is identified which 
makes the contribution of the jumps equal to zero after
the contraction with the real photon polarization vector is performed.

The divergencies in the amplitude are
 related to the fact that the additional functions obtained
with the
help of WW-like relations generically contain discontinuities at
the points $x=\pm \xi$ what in turn leads to formally divergent results for
the DVCS amplitude with a transversely polarized photon. However
the divergencies contribute to the observables only at the
accuracy $1/Q^2$ and can be neglected at twist-3 accuracy.
We observed that the divergencies are canceled in the twist-3
DVCS amplitude with a longitudinally polarized virtual photons.
Such a cancelation of the divergencies in the longitudinal twist-3
amplitude implies that the DVCS observables up to accuracy
$O(1/Q)$ can be estimated using the WW approximation, at least
at leading order in $\alpha_s$. The case of DVCS off the nucleon will be
studied elsewhere.

\section*{\normalsize \bf Acknowlegments.}

We would like to thank  I.~Anikin, V.~Braun, A.~Belitsky,
M.~Diehl, L.~Frankfurt, K.~Goeke, L.~Mankiewicz, D.~M\"uller,
M.~Penttinen, P.~Pobylitsa, M.~Strikman for conversations.
Recently we learned that the divergencies of twist-3 DVCS
amplitude in WW approximation were independently found by
A.~Radyushkin and C.~Weiss \cite{RW}. We are grateful to them for
the discussions. The work of N.K. was supported by the DFG,
project No. 920585.
O.V.T. was partially supported by RFFI grant 00-02-16696.
 N.K. and O.V.T. are thankful to Klaus Goeke
for invitation to Bochum University where the idea to write these
notes has appeared.

\section*{\normalsize \bf APPENDIX: Derivation of the WW relations}
\setcounter{equation}{0}
\label{app:a}
\renewcommand{\theequation}{A.\arabic{equation}}
\setcounter{table}{0}
\renewcommand{\thetable}{\Alph{table}}

Here we briefly  describe the method used to derive  the WW
relations for skewed parton distributions.
Our approach is very close to approach
used in \ci{BM}. The difference lies only in  technical details.
We have also been informed that  similar results have been
obtained independenly in \ci{RW}.

Our starting point  is the following equations derived in \ci{BB89}:
\be{vect}
\pbar (x)\gamma_\mu \p (-x) &=&\int_0^1\, dt\,
{ \partial \over \partial x_\mu}\,\pbar(tx)\xslash
\p (-tx)-
i\epsilon_{\mu\nu\alpha\beta}\int_0^1\,dt\,t\,x^\nu\pert^\alpha
\left[ \pbar(tx)\gamma^\beta\gamma_5\p (-tx)\right]-
\nonumber\\
&&\mskip-50mu{}-
 \int_0^1\,dt\,\int_{-t}^t\,dv\,\pbar(tx)
\left\{\,
v\,igG_{\mu\nu}(vx)-t\,\gamma_5 g\Gtilde_{\mu\nu}(vx)
\right\}
x^\nu \xslash \p(-tx)
\ee
and
\be{avect}
\mskip-10mu
\pbar (x)\gamma_\mu\gamma_5 \p (-x) &=&\int_0^1dt
{ \partial \over \partial x_\mu}\,\pbar(tx)\xslash \gamma_5
\p (-tx)- i\epsilon_{\mu\nu\alpha\beta}\int_0^1dtt\,x^\nu\pert^\alpha
\left[ \pbar(tx)\gamma^\beta  \p (-tx)\right]
\nonumber\\
& &\mskip-50mu{}- \int_0^1\,dt\,\int_{-t}^t\,dv\,\pbar(tx)
\left\{
t\, g\Gtilde_{\mu\nu}(vx)+ v\,\gamma_5 igG_{\mu\nu}(vx)
\right\}
x^\nu \xslash \p (-tx)\nonumber\\
\ee
where we do not write out explicitly  all the path-ordered gauge factors,
and  $\pert_\alpha$ denotes the total
derivative defined as :
\be{Dtot}
\pert_{\alpha}\left\{ \pbar(tx)
\Gamma [tx, -tx] \p(-tx) \right\} \equiv
\left. \frac{\partial}{\partial y^{\alpha}}
\left\{ \pbar(tx + y) \Gamma
[tx + y, -tx + y] \p(-tx + y)\right\} \right|_{y \rightarrow 0},
\ee
with generic Dirac matrix structure $\Gamma$ and
$[x,y] =\mbox{\rm Pexp}[ig\!\!
\int_0^1\!\! dt\,(x-y)_\mu A^\mu(tx+(1-t)y)]
$.
Note that in the matrix elements  the total
derivative can be easily converted to  the momentum transfer:
\be{exmpl}
\langle p'|\pert_\mu
\pbar(tx)\Gamma [tx, -tx] \p(-tx) |p\rangle=i(p'-p)_\mu
\langle p'|\pbar(tx)\Gamma [tx, -tx] \p(-tx) |p\rangle
\ee

The general method to obtain Wandzura-Wilczek relations is
to take the matrix elements in the LHS and RHS
\re{vect} and \re{avect} and insert their  parametrisation.
This provides us
with a system of equations for the twist three functions like
$H_3$ and $H_A$. Such a method was effectively also used for
the $\rho$-meson distribution amplitudes, see \ci{BB89,BBKT}.
 In the case of  skewed distributions it is
more convenient to solve  the system of equations \re{vect},
\re{avect} at the operator level and then take the matrix elements.
Operator solution means that one has to express
{\it non-symmmetrical} operator
$\pbar (x)\gamma_\mu\gamma_5 \p (-x)$ through the
two point {\it symmetrical}
operators $\pbar (x)\xslash(\gamma_5) \p (-x)$ and three
point quark-gluon operators.

Consider as an example the solution for the vector operator.
Substituting \re{avect} into \re{vect} we obtain an equation
which contains only  one non-symmetrical  vector operator :
\be{reduce}
u\,\pbar (ux)\gamma_\mu \p (-ux)&=&
[(x\pert)^2-x^2(\pert)^2]\int_0^u\,dt\,t(u-t)
\pbar (tx)\gamma_\mu \p (-tx)+
\nonumber\\
&&\mskip-50mu
 + { \partial \over \partial x_\mu}\, \int_0^u\, dt\,
\pbar(tx)\xslash \p (-tx)+
[x_\mu\pert^2-(x\pert)\pert_\mu] \int_0^u\, dt\,\,t(u-t)
\pbar (tx)\xslash\p (-tx)
\nonumber\\
&&\mskip-50mu
- i\epsilon_{\mu i j k}x^i\pert^j{ \partial \over \partial x^k}
\int_0^u dt\, (u-t)
\left[ \pbar(tx)\xslash\gamma_5  \p (-tx)\right] +\, \dots
\ee
where ellipses stands for the contributions of the
three point quark-gluon operators. It is convenient to rewrite this
equation in the compact form:
\be{intEQ}
f(u)= k^2\int_0^u (u-t)f(t) + \varphi(u)
\ee
where we introduced the following notations:
\be{Snot}
f(u)&=& u\,\pbar (ux)\gamma_\mu \p (-ux), \quad f(0)=0,
\nonumber\\
k&=& \sqrt{(x\pert)^2-x^2(\pert)^2}
\ee
and $\varphi(u)$ denotes the  known terms in the RHS of the \re{reduce}.
Our aim is to solve this integral equation and find $f(u)$.
It it easy to see that \re{intEQ} can be reduced  to the simple
second order differential equation:
\be{difEQ}
 f''(u)= k^2\,f(u) + \varphi''(u),
\ee
with the boundary conditions:
\be{Bcon}
f(0)=0,\quad
f'(0)= {\partial \over \partial x_\mu} \pbar(0)\xslash \p(0)
\ee
The solution is
\be{Soltn}
f(t)=e^{-tk}\int_0^t\, du e^{2uk}\left[
f'(0)+\int_0^u d\alpha e^{-\alpha k}\varphi''(\alpha)
\right]
\ee
Substituting the definitions \re{Snot} we obtain an explicit expression.
We then simplify the full answer by neglecting  the square
of the momentum transfer. This corresponds to the formal substitutions
$ \pert^2=0, \quad k=x\pert$.
Then after simple manipulations we obtain the following operator
equation:
\be{reslt}
\pbar ( x)\gamma_\mu \p (- x)=\mskip300mu
\nonumber\\
 \frac12 \int_0^1 d\alpha
\left\{
\alpha\,   \pert_\mu
\left(
e^{-  \bar \alpha (x\pert)}-e^{  \bar \alpha (x\pert)}
\right)
+\left(
e^{-  \bar \alpha (x\pert)}+e^{  \bar \alpha (x\pert)}
\right){\partial \over \partial x_\mu}
\right\}
\pbar( \alpha x)\xslash \p(- \alpha x)
\nonumber\\
 - i\epsilon_{\mu i j k}  x^i\pert^j
\int_0^1 du\, e^{(2u-1) (x\pert)}
\int_0^u d\alpha e^{-\alpha   (x\pert) }
{ \partial \over \partial x_k}
 \pbar( \alpha x)\xslash\gamma_5  \p (- \alpha x) +\, \dots
\mskip60mu
\ee
where $\bar\alpha=1-\alpha$ and ellipses are contributions of the three
point quark gluon operators which we do not write explicitly for the
sake of simplicity. This operator relation can be used for the derivation
of various WW relations, because in the RHS of the eq.~(\ref{reslt})
only two-point quark operators of twist-2 contribute in this case.
Also the general solution \re{Soltn} can be applied for all-order
summation of the kinematical power suppressed terms in the DVCS
amplitude.

\end{document}